%% file: main.tex
\documentclass[a4paper, amsfonts, amssymb, amsmath, reprint, showkeys, nofootinbib, twoside]{revtex4-1}
\usepackage[english]{babel}
\usepackage[utf8]{inputenc}
\usepackage[colorinlistoftodos, color=green!40, prependcaption]{todonotes}
\input{preamble}
\usepackage[pdftex, pdftitle={Article}, pdfauthor={Author}]{hyperref} 
\begin{document}
\title{Dynamical analysis of quantum matter bounces with dark sector mimickers}


\author{Francisco  Bento Lustosa}
\email[Correspondence email address: ]{chicolustosa@cbpf.br}
\author{Nelson Pinto-Neto}
    \affiliation{Centro Brasileiro de Pesquisas Físicas,
Rua Dr. Xavier Sigaud 150, Urca, CEP: 22290-180, Rio de Janeiro-RJ, Brazil}

\date{\today} 

\begin{abstract}
We study the effects of the inclusion of fluids 
in bounce scenarios driven by an exponential potential scalar field. Most solutions exhibit well known tracking behavior between the fluids and the scalar field. This tracking behavior can model transitions between different phases of cosmic evolution. We will focus on an interesting bouncing model with a dust matter fluid, where the scalar field can drive an early dark energy expanding period with a radiation-like dominated phase just after it, and then tracks the dust matter fluid with energy density compatible with the dark matter energy density. The model is dust dominated in the far past of the contracting phase, and has stiff matter behavior when approaching the singularity, allowing well known quantum bounce transitions to the expanding era. Hence, it is a quantum matter bounce scenario with an inflationary phase together with a smooth transition through a radiation era to matter domination with a possible scalar field dark matter candidate.
\end{abstract}


\maketitle

\input{sections/section01.tex}  
\input{sections/section02.tex}
\input{sections/section03.tex}
\input{sections/section04.tex}
\input{sections/section05.tex}
\input{sections/conclusions}

\bibliography{bibliography.bib}

\end{document}

%% file: preamble.tex
\usepackage{amsthm}
\usepackage{float}
\usepackage{mathtools}
\usepackage{physics}
\usepackage{xcolor}
\usepackage{graphicx}
\usepackage[margin=.5in]{geometry} 
\usepackage{adjustbox}
\usepackage{placeins}
\usepackage[T1]{fontenc}
\usepackage{lipsum}
\usepackage{csquotes}
\usepackage{caption}
\usepackage{color}
\usepackage{soul}

%% file: sections/section01.tex
\section{INTRODUCTION} \label{sec:outline}

The golden age of precision cosmology continues to provide us with observations that challenge the standard inflationary $\Lambda$CDM model. Starting with the observation that our universe is currently undergoing an accelerated expansion \cite{SupernovaSearchTeam:1998fmf,Astier2012darkenergyobservations}, which reaffirmed the cosmological constant problem \cite{Martin2012cosmologicalconstantproblem}, and up to the current discussions surrounding the Hubble and $S_8$ tensions \cite{DIVALENTINO2021hubble, DIVALENTINO2021s8}, there are several aspects of the standard model that are being scrutinized and debated. In addition to that, one of the main features of the standard cosmological model, cold dark matter, has yet to be detected in high energy experiments providing strong bounds on its possible particle nature. 

A core aspect of the inflationary model that is in accordance with observations is the production of an almost scale invariant spectrum for scalar perturbations \cite{Planck2018inflation}. In particular, using a scalar field with an exponential potential, one can model power-law inflation and some other interesting models with positive or negative potentials. There are many interesting features when one analyzes the cosmological evolution of an expanding Friedmann universe driven by such scalar fields alone \cite{Heard_2002}. For models starting in a Big Bang type singularity, it has been shown that the scalar field behaves as stiff matter ($p=\rho$) initially and evolves into a potential-kinetic scaling behavior at late times with constant $w_{\phi}$ ($p = w_{\phi}\rho$). The solutions can also pass through a transient phase where $w_{\phi} = -1$ that can be seen either as an inflationary phase or a dark energy phase, depending on the scales and parameters used.The addition of perfect fluids to this simple model offers a plethora of different cosmological models, whose classical aspects have been analyzed in detail in Refs.~\cite{Copeland1998,Heard_2002}.

Setting aside the current observations putting the $\Lambda$CDM model under debate, there are other conceptual issues that are present in inflationary scenarios that have not been solved \cite{bib:Martin2001,brandenberger2012} and are arguably beyond the - current - scope of the Standard Model of both Cosmology and Particle physics. The singularity problem, the fine tuning in the initial conditions necessary to yield the correct amplitude of perturbations and the energy scale where inflation takes place point to the necessity to consider physics beyond effective field theory models \cite{Brandenberger2022}. An alternative scenario that solves or avoids some of this issues is the matter bounce scenario that has gained increased attention in the literature in the past 20 years \cite{Finelli2002, Wands2004, Novello_2008, Peter_2008, brandenberger2012, Brandenberger2012b, Cai_2012, Brandenberger_2017}. In this type of scenario there is no horizon or flatness problem due to the long phase of contraction before the bounce. The challenge is to provide a mechanism that generates the bounce and avoids issues regarding instabilities of perturbations \cite{Belinsky1970, Battefeld2014, Brandenberger_2017, Vitenti2012, Pinto-NetoVitenti2013} and trans-Planckian physics \cite{bib:Martin2001, brandenberger2012, bib:Valentini2014}. 

The exploration of cosmological models featuring a bounce necessitates the incorporation of novel physics, whether through matter fields that violate energy conditions or through modifications to General Relativity, potentially inspired by quantum gravity considerations, see, e.g. \cite{Brandenberger2020},  Another example is Ref.~\cite{Wands2004}, where the authors were obliged to substitute the additional perfect fluid appearing in Refs.~\cite{Copeland1998,Heard_2002} for a negative energy scalar field in order to obtain a bounce. However, if the bounce arises due to quantum gravitational effects, there is no need to call on negative energy fields, and one can stay with the original analysis developed in Ref.~\cite{Copeland1998,Heard_2002}. The aim of this paper is to study models with a quantum bounce in the context of Ref.~\cite{Heard_2002}, and select viable interesting possibilities which deserves to be further studied.

Quantum gravitational effects in Cosmology have ben studied in the context of Loop Quantum Cosmology \cite{Wilson-Ewing2012,deHaro2014,Angullo2022}, ekpyrotic models, and the Wheeler-DeWitt approach. Our analysis and discussions remain largely invariant under different assumptions about the scalar field as long as its kinetic term is canonical. We also point out that the evolution of perturbations from the contracting phase through the bounce was a subject of contention in the literature for many years before being settled for a class of models in \cite{Vitenti2012, Pinto-NetoVitenti2013} that includes the case of the exponential potential scalar field. It was shown in \cite{Vitenti2012, Pinto-NetoVitenti2013} that assuming an initially homogeneous and isotropic Universe, where inhomogeneous cosmological perturbations are present only as quantum vacuum fluctuations, then shear perturbations will not become nonlinear up to the bounce if it is not very deep, with curvature scale of the order of the Planck length, and the known Belinsky-Khalatnikov-Lifshitz instabilities will not be present.

Quantum cosmology has not only been used to study proposals for a complete cosmological model, but it is also a fertile ground for the application of different quantum theories and interpretations, such as the Consistent Histories approach \cite{hartle2014spacetimequantummechanicsquantum, Anastopoulos_2005, Halliwell2006} or the Many Worlds interpretations of Quantum Mechanics \cite{DeWitt1973-DEWTMI}. That is due to the fact that in quantum cosmology the measurement problem becomes more explicit due to the lack of observers in the early universe. Besides that, the quantum-to-classical transition through decoherence of cosmological perturbations is still a subject of debate in the literature \cite{bib:Sudarski2016}. 

Another possibility that has been successfully applied to quantum cosmology for the construction of a number of interesting models is the de Broglie-Bohm or Pilot-Wave Theory (PWT) \cite{bib:Pinto-Neto2021}. In this theory all systems, including the universe, are described by a wavefunction or wavefunctional together with their actual configuration that evolves according to de Broglie's guidance equations. In these equations, a natural measure of the quantum effects emerge, the quantum potential, which deviates the quantum trajectories from the classical ones and bring in it the strange features of quantum mechanics, as non-locality. In quantum cosmology, the wave-functional is obtained as a solution of an appropriate quantum gravity functional differential equation, the simplest and more straightforward one being the Wheeler-DeWitt equation (which is however viewed as an effective equation applicable to very high energy scales but not the Planck energy itself), once an initial configuration is assumed. In the PWT approach, the possible evolutions for the background and perturbations can be obtained from appropriate guidance equations. The classical limit is naturally obtained by analyzing the quantum potential and inspecting when it becomes negligible with respect to the classical terms of the guidance equations. A quantum bounce can occur due to quantum gravitational effects that can be effectively modeled by the quantum potential. If the quantum bounce does not happen very near the Planck length, then one can rely on the simplest approach to quantum cosmology: the Wheeler-DeWitt quantization. Pilot-wave theory has also been used in the quantization of fields through different approaches \cite{Struyve_2011, Fabbri2022, Nikolic2022} and although Lorentz invariance is broken at a fundamental level it can be recovered as an emergent symmetry and is always respected when it comes to observers in quantum equilibrium\footnote{In Pilot-wave theory the Born rule is not postulated and is hypothesized that arbitrary initial configurations will, in general, evolve to configurations given by $\rho = |\psi|^2$. This is the Quantum Equilibrium Hypothesis (QEH) and it has been corroborated by various numerical simulations that show that sufficiently complex quantum systems that initially violate the Born rule (hence, are out of equilibrium) relax in very short time scales to the equilibrium distribution. For more details and references see \cite{bib:Lustosa2020, Lustosa2023}.} \cite{valentini2024brogliebohmquantummechanics}. There are still open questions regarding how to define probabilities for the quantum state of the universe \cite{Valentini2023}, but the theory can be used as an effective tool for deriving predictions in quantum cosmological models, and can even be used in the context of Loop Quantum Cosmology \cite{Sen_2024}. Furthermore, in the framework of PWT a simple solution for the quantum-to-classical transition of cosmological perturbations has been obtained \cite{bib:Pinto-Neto2012}.

In recent works, making use of the results presented in Ref.~\cite{Heard_2002}, one of the authors has developed a scalar field matter bounce model \cite{colinpinto-neto2017, Bacalhau2018, Frion2024bounce} that generates an almost scale invariant spectrum of scalar perturbations, and a ratio between scalar and tensor amplitudes in accordance with recent observations \cite{Tristram2022tensor-to-scalarratio}. The model has a contracting phase dominated by dust, passes through a quantum bounce described by pilot-wave quantum cosmology \cite{bib:Pinto-Neto2021}, and goes into an expanding phase with a period of transient accelerated expansion occurring before the scalar field recovers the dust-like behavior. The long dust dominated contraction allows for the generation of the scale invariant scalar perturbations and the quantum bounce amplifies them over the tensor perturbations generating the ratio $r \lesssim 0.1$. The quantum cosmological model provided in the context of the de Broglie-Bohm theory has the distinct advantage of providing a continuous description of the transition between classical and quantum phases, both for the background and the perturbation variables. As a side remark, one of the main features of the model studied in \cite{colinpinto-neto2017, Bacalhau2018, Frion2024bounce} was the fact that the scalar field can play the role of dark energy in the expanding phase, as the almost scale-invariant spectrum of scalar perturbation was generated by the long phase of contraction dominated by the same scalar field which acts as dust in this era. As the scalar field behaves as dark energy only in the expanding phase, this model is not pervaded by an issue that arises when considering bouncing models with a cosmological constant, which is to define adiabatic quantum vacuum states for the perturbations in the far past when it is dominated by the cosmological constant \cite{Pinto-Neto2012cosmologicalconstant}.

We will use the model with a single scalar field with an exponential potential described above as a starting point for our analysis. As we will show, in the presence of an additional fluid classically discussed in Ref.~\cite{Heard_2002}, the contraction (expansion) phase will generally end (start) with the kinetic term of the scalar fluid dominating over the matter and potential components \cite{Wands2004}. This will allow us to build our models assuming a quantum bounce fully dominated by the kinetic energy of the scalar field, and described by the quantum trajectories already obtained in Ref.~\cite{Pinto-Neto2000}. Such a bounce allows as to connect the variety of possibilities of singular contracting or expanding cosmological models presented in Ref.~\cite{Heard_2002}, allowing the construction of many different quantum bounce scenarios. Although not exhaustive, in our analysis we present some representative solutions of some sets of models, and the most physically interesting ones.

This work is organized as follows. In Section \ref{sec:develop} we apply the phase space analysis of \cite{Copeland1998,Heard_2002} to the quantum bounce scenario, discussing the general properties of the classical limits for the expanding and contracting phases. In Section \ref{sec:quantumbounce} we describe how a quantum bounce that leads to the classical phase space equations of motion can be constructed in the context of the de Broglie-Bohm theory. In Section \ref{sec:phasespace} we will provide our analysis of some representative models of different sets of scenarios, and the most physically appealing ones. In Section V we comment our results, and we discuss prospects for future work.


%% file: sections/section02.tex
\section{Phase space dynamics for the matter bounce scenario} \label{sec:develop}

In \cite{Heard_2002} the phase space analysis of the two-dimensional phase space equations we will study in this work was focused on possible expanding scenarios and their future infinity features. In the final paragraph of \cite{Heard_2002}, Heard and Wands mention that a collapsing model giving a scale-invariant spectrum of curvature perturbations \cite{Finelli2002} indeed corresponds to a kinetic-potential unstable solution that leads to a kinetic-dominated collapse. In the next subsection we expand on that point, describing qualitatively the behavior of the solutions for the case with only one scalar field with an exponential potential, reviewing the results of Ref.~\cite{Bacalhau2018}. In subsection B we extend the analysis to a 2-dimensional phase space corresponding to the scalar field and an additional barotropic fluid. We make use of the symmetries between collapsing and expanding solutions already mentioned in \cite{Heard_2002} so these section is a review of their analysis and  with a focus on bouncing models. The fact that we recover the qualitative behavior of their solutions in phase space in Section \ref{sec:phasespace} serve as a check on the validity of our following numerical analysis.  

\subsection{One-dimensional phase space}

Consider a spatially-flat Friedmann-Lemaître-Robertson-Walker (FLRW) universe which contains only a scalar field with an exponential potential energy density. Defining our exponential potential as
\begin{equation}
    V = V_0 e^{-\lambda \kappa \phi},
\end{equation}
where $\kappa \equiv 8\pi G_N$, we can write the Klein-Gordon equation in the expanding background 
\begin{equation}
    \ddot{\phi} + 3 H \dot{\phi} + \frac{d V}{d \phi} = 0.
\end{equation}
The Hubble parameter is determined by the Friedmann constraint 
\begin{equation}
H^{2} = \frac{\kappa^{2}}{3}\left ( \frac{\dot{\phi}^{2}}{2} + V\right ),
\end{equation}
and the scalar field has the usual energy density and pressure defined, respectively, as $\rho_{\phi} = \dot{\phi}^2/2 + V$ and $P_{\phi} = \dot{\phi}^2/2 - V$. Their ratio will determine the equation of state (EoS) parameter for the field at any time
\begin{equation}
    \frac{P_{\phi}}{\rho_{\phi}} = w_{\phi}.
\end{equation}
Considering only $V > 0$ the  EoS parameter can range from a kinetic-dominated solution where $w_{\phi} = 1$ to a potential dominated solution where $w_{\phi} = -1$. To determine how the field will actually evolve and how it will affect the expansion of the universe we have to provide initial conditions to solve equation (2). Since it is a second order equation we need both $\dot{\phi}(0)$ and $\phi(0)$. Following \cite{Copeland1998, Heard_2002} we will use the Friedmann constraint to define a new set of dimensionless variables
\begin{equation}
    1 = \frac{\kappa^{2}}{3}\left ( \frac{\dot{\phi}^{2}}{2H^{2}} + \frac{V}{H^{2}}\right) \equiv (x^2 + y^2).
\end{equation}
The EoS parameter for the scalar field is now defined as 
\begin{equation}
   w_{\phi} = \frac{x^{2} - y^{2}}{x^{2} + y^{2}},
\end{equation}
and the Klein-Gordon equation can be recast into an autonomous system:
\begin{align}
 {x}' = -3 x \left ( 1 - x^{2} \right) \pm \lambda \sqrt{\frac{3}{2}}y^{2}, \\
 {y}' = x y \left ( 3 x - \lambda\sqrt{\frac{3}{2}} \right),
\end{align}
where a prime denotes derivation with respect to $\alpha = \ln (a)$. As the system is subject to the constraint (5) we have a one-dimensional phase space of solutions on the unit circle. 

\subsubsection{Critical points}
\begin{figure}[H]
\centering
\includegraphics[scale=0.45]{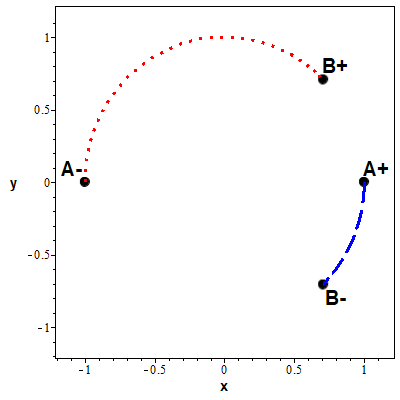} 
    \caption{Phase space trajectory of a contracting (dashed blue) and expanding (dotted red) universe with a standard scalar field with potential  $V = \exp{-\sqrt{3}\kappa \phi}$ ($\lambda = \sqrt{3}$). At the points $B\pm$ the field behaves as dust $w_{\phi} = 0$. The quantum bounce connects $A+$ to $A-$, where the field behaves as stiff matter ($w_{\phi} = 1$). In the expansion phase when the trajectory goes through the point $y=1$ there is an accelerated expansion phase with $w_{\phi} = -1$.}
    \label{fig:phasespace1}
\end{figure}

Critical points correspond to fixed points where ${x}'=0, {y}'=0$. In those points we have 
\begin{equation}
    \frac{\dot{H}}{H^2} = -3 x_i^2 = C,
\end{equation}
where $C$ is an integration constant. All critical points where $x_i$ is a non-zero constant correspond to power-law solutions for the scale factor as a function of cosmic time: $a \propto |t|^p$, with $p = 1/(3x_i^2)$. The one-dimensional autonomous system has three fixed points (Figure \ref{fig:phasespace1}):
\begin{itemize}
    \item $A_{\pm}$ : kinetic-dominated solutions with $y_A=0$ and $x_{A_{\pm}} = \pm 1$. In this case $a \propto t^{1/3}$ and we have a stiff-matter type evolution. 
    
    \item $B$ : potential-kinetic-scaling exists for $\left(6-\lambda^{2} \right) > 0$ and happens when $x_B = \frac{\lambda}{\sqrt{6}}, y_B = \sqrt{1- \frac{\lambda^{2}}{6}}$. In this case $a \propto t^{2/\lambda^2}$ and for $\lambda^2 < 2$ it correspond to power-law inflation. Near this point the effective EoS parameter becomes $$w_{\phi} = (\lambda^2 -3)/3 \; .$$ In Section IV we will consider two physically interesting cases: $\lambda = \sqrt{3}$ and $\lambda = 2$, corresponding to dust and radiation, respectively.
\end{itemize}
\subsubsection{Stability}

\begin{itemize}
    \item $A_{\pm}$: $x_{A_{\pm}} = \pm 1$ are always unstable points about linear perturbations of the form $x \rightarrow x + u$ when $\lambda^2 < 6$ and positive. The perturbations have exponential behavior with $u \propto e^{m_{\pm}N}$, where $m_{\pm} = \sqrt(6)(\sqrt(6) \mp \lambda)$.
    \item B: Linear perturbations always decays when this point exists for positive potentials, so the solution is always stable. 
\end{itemize}

\begin{figure}[H]
\begin{tabular}{c}
\includegraphics[width=.5\textwidth, height=1.2in]{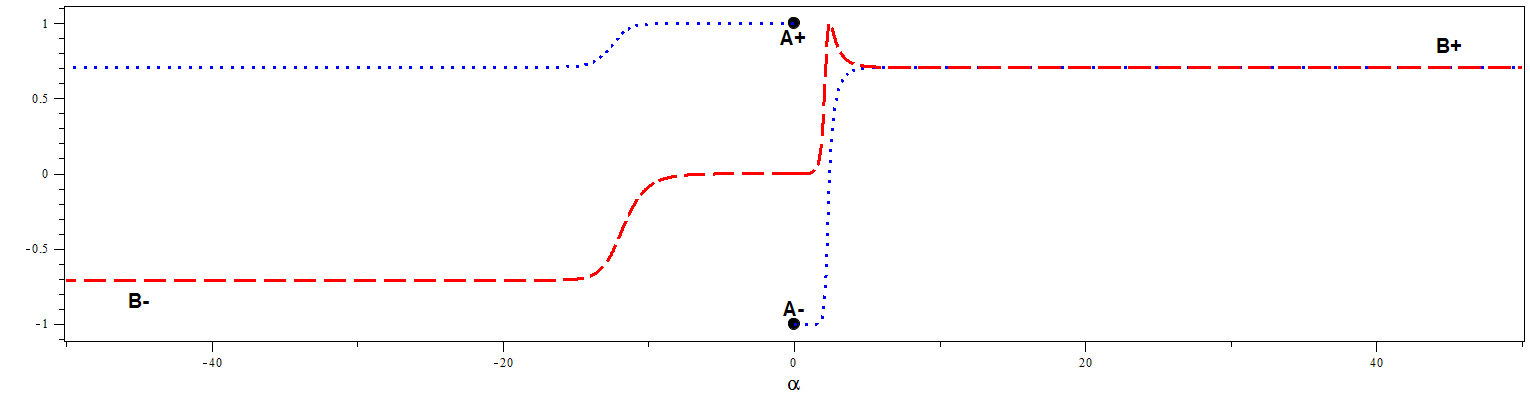}  \\ (a)  \\ 
  \includegraphics[width=.5\textwidth, height=1.2in]{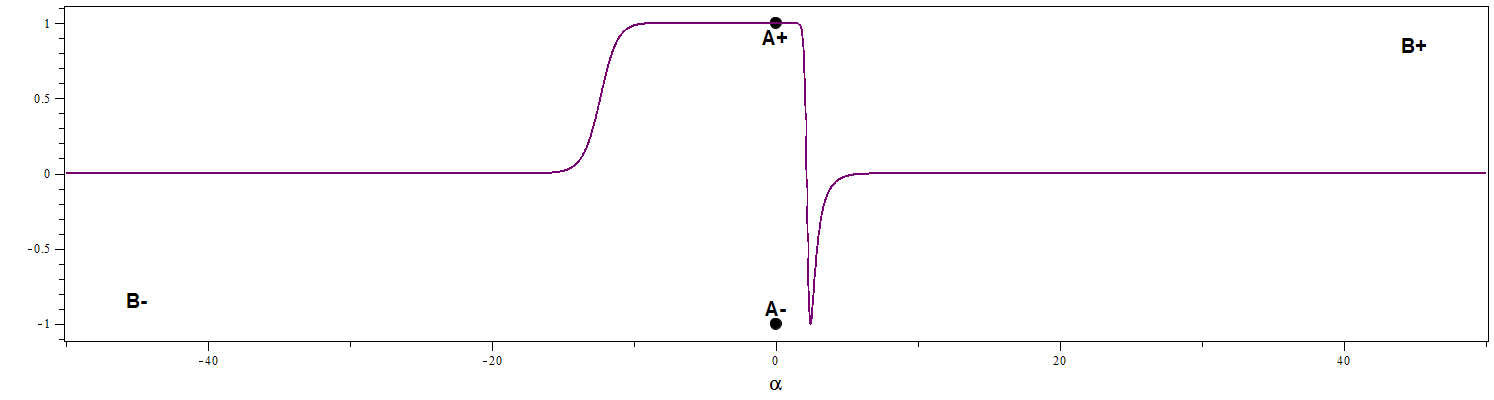} \\
(b) \\
\end{tabular}
\caption{Evolution of the variables $x(\alpha)$ (dotted blue) corresponding to the kinetic term and $y(\alpha)$ (red dashed) corresponding to the potential term of Figure \ref{fig:phasespace1} (a) and the EoS parameter for the scalar field $w_{\phi}$ (b) with respect to $\alpha = \ln{a}$.}
    \label{fig:xywevol1}
\end{figure}

For a description of late time behavior and qualitative evolution for negative and/or steep potentials we direct the reader to reference \cite{Heard_2002}. The case were $V >0$ and $\lambda = \sqrt{3}$ was the one studied in \cite{colinpinto-neto2017, Bacalhau2018, Frion2024bounce} so we use it as our first example for our analysis in Figures \ref{fig:phasespace1} and \ref{fig:xywevol1}. In this case the field has a dust-like equation of state in the far past and the far future, starting at the critical point $B_{-}$ and contracting towards $A_{+}$ where the field becomes dominated by the kinetic term and the equation of state (EoS) parameter becomes $w_{\phi}=1$. At this point, we assume that the quantum bounce takes place and the singularity can be avoided by quantum gravitational effects that can be effectively described by the quantum potential appearing in the Pilot-wave theory of canonical quantum gravity \cite{Pinto-Neto2000,colinpinto-neto2017, Bacalhau2018, Frion2024bounce}. We will briefly describe how this happens in Section \ref{sec:quantumbounce}, but the bounce serves essentially to connect the points $A_{\pm}$ to $A_{\mp}$ in a continuous way that leads to an initial state of expansion still dominated by the kinetic term of the scalar field. When $\lambda = \sqrt{3}$ after the stiff matter phase the field becomes dominated by the exponential potential for a brief period before returning to a scaling behavior that makes it dust-like once more. In Figure \ref{fig:xywevol1} we can see the bump in the $y(\alpha)$ red dashed line right after the bounce which, depending on the parameters of the model and the corresponding time scales, can be interpreted either as an early inflationary expansion, or the present day dark energy accelerated phase \cite{Bacalhau2018}. In either case the evolution can be qualitatively described by the phase space trajectories of Figure \ref{fig:phasespace1}, and further analyses by viewing how each degree of freedom evolves individually with time. We also plot the evolution of the EoS parameter of the scalar field as we will be interested in how it affects the overall evolution and how it compares to the EoS parameter of the other fluids we will add to our model. 

\subsection{Two-dimensional phase space}

Consider now the addition of an extra component of matter described by a fluid with energy density $\rho_m$, which does not interact with the scalar field. Hence, it obeys a separate continuity equation of the standard form 
\begin{equation}
    \dot{\rho}_{m}= 3 H \left ( \rho_{m} + P_{m} \right) \; . 
\end{equation} 
We assume that the fluid obeys a barotropic equation of state $P_m = w \rho_m$, and we will work with the standard values: $w = -1$ (dark energy), $w = 0$ (dust), $w = 1/3$ (radiation). 
The Friedmann constraint now reads
\begin{equation}
        H^{2} = \frac{\kappa^{2}}{3}\left ( \frac{\dot{\phi}^{2}}{2} + V + \rho_{m} \right ) \; .
\end{equation}

Defining a new variable for the matter component as
\begin{equation}
        z \equiv \frac{\kappa \sqrt{\rho_{m}}}{\sqrt{3}H}, 
\end{equation}
we can rewrite the Friedmann constraint as 
\begin{equation}
        x^{2} \pm y^{2} + z^{2} = 1
\end{equation}
and obtain the dynamical equations given by \cite{Heard_2002}:
\begin{align}
    {x}' = -3 x \left ( 1 - x^{2} - \frac{\gamma}{2}z^{2}\right) \pm \lambda \sqrt{\frac{3}{2}}y^{2}, \\
   {y}' = y \left ( 3 x^{2} + \frac{3\gamma}{2}z^{2} - \lambda\sqrt{\frac{3}{2}}x \right), \\
    {z}' = \frac{3}{2}z \left ( -\gamma + 2 x^{2} + \gamma z^{2} \right),
\end{align}
where $\gamma = w + 1$.
\subsubsection{Critical points}
Once more, critical points correspond to fixed points where ${x}'=0, {y}'=0, {z}'=0$. The two-dimensional system can have up to five critical points, three given by the scalar field solutions of the one-dimensional system $A_{\pm}$, $B$ and two given by:
\begin{itemize}
    \item C : A fluid-dominated solution where $z_c = 1$ always exists and corresponds to $a \propto |t|^{2/3\gamma}$; 
    \item D : A scaling solution between all three components where 
    \begin{equation}
        x_D = \sqrt{\frac{3}{2}}\frac{\gamma}{\lambda}, \ y_D = \sqrt{\pm \frac{3}{2}\frac{(2-\gamma)\gamma}{\lambda^{2}}}, \ z_D = 1 - \frac{3\gamma}{\lambda^{2}}.
    \end{equation}
    This solution only exists when $\lambda^2 > 3\gamma^2.$ The scale factor behaves exactly like in the fluid-dominated solution and is independent on the slope of the scalar potential. 
\end{itemize}

\subsubsection{Stability}

When fluid perturbations are introduced, the stability depends on a second eigenmode beyond $u$, given by $z = 3(2 x_i^2 - \gamma)z/2$.
\begin{itemize}
    \item $A_{\pm}$: The perturbations are also exponentially unstable with eigenvalue given by $m=3(2-\gamma)/2$ (except for the case where $\gamma = 2$, but we will only be considering cases where $\gamma < 2$). 
    \item B: In this case, solutions are stable when $\lambda^2 < 3\gamma$ and unstable when $\lambda^2> 3\gamma$. It is interesting to note that when $\lambda = 2$ and $\gamma = 4/3$ the fluid is radiation-like and the solutions are marginally stable in this case. 
    \item C: For generic perturbations this is a saddle point when $0 < \gamma < 2$, so it is always unstable in the cases of interest. This is an important point to note, since it implies the necessity of the presence of the scalar field at all times if we want the model to become stable. 
    \item D: This solution, when it exists, is always the late-time stable attractor. 
\end{itemize}

We again refer the reader to reference \cite{Heard_2002} for further details on the general analysis of the classical behavior. Using their language, in this work we will be interested only in scalar fields with positive \textit{flat} potentials that lead to tracking or scaling solutions at early and late times in our models. A detailed analysis of the case with $\lambda = \sqrt{3}$ and $\lambda = 2$ with different standard barotropic fluids will be given in Section \ref{sec:phasespace}. Before going into that it is worthwhile to briefly describe how the bounce appears using the Pilot-Wave theory. It serves to justify the connection between the critical points $A_{\pm}$,  when the model evolution is dominated by the scalar field kinetic energy, allowing direct application of the results of Ref.~\cite{Pinto-Neto2000}. Note that whenever a cosmological model contains a canonical scalar field with general potentials, including the majority of inflationary potentials, and all other fluids satisfy $P/\rho < 1$, then it is the kinetic energy density of the scalar field which dominates near the singularity. Hence, the bounce solution of Ref.~\cite{Pinto-Neto2000} applies to a wide class of cosmological models, without the need of any exotic potential or curvature couplings.

%% file: sections/section03.tex
\section{The Quantum Bounce} \label{sec:quantumbounce}

The General Theory of Relativity has its limits of application most clearly demonstrated when it concerns the singularity problem. Although the standard inflationary $\Lambda$CDM model assumes the Universe started in a singularity (or at least the model points to this conclusion), there is a growing number of authors that has accepted that some form of quantum gravitational effect might prevent the formation of singularities. At this point, we do not have a complete theory of quantum gravity due to the nonrenormalizable character of General Relativity when considered as a field theory and also due to other conceptual issues related to the foundations of quantum mechanics \cite{bib:Pinto-Neto2013}. Quantum foundations and quantum cosmology are naturally connected by the measurement problem and the interpretation of the wavefunction, since when we consider the universe as a whole it becomes impossible to talk about observers and measurements being the cause for collapse or decoherence. This is quite an old discussion and there are many proposed quantum theories that circumvent this issue like the many-worlds interpretation \cite{DeWitt1973-DEWTMI}, different history-based approaches \cite{hartle2014spacetimequantummechanicsquantum, Anastopoulos_2005, Halliwell2006} and the de Broglie-Bohm or Pilot-Wave theory \cite{bib:Pinto-Neto2021}, that we will use here to describe the quantum bounce in our model \cite{colinpinto-neto2017, Bacalhau2018, Frion2024bounce}. 

The Pilot-Wave Theory assumes that each degree of freedom of a given system has a trajectory in configuration space guided by the wavefunction of the whole system. This allows for the trajectories to be non-locally correlated although not signaling when systems are in quantum equilibrium (i.e. obeying the Born rule) \cite{bib:Lustosa2020, Lustosa2023}. Hence, explicit Lorentz symmetry violation cannot be detected in experiments when the Born rule holds. In ``standard'' quantum cosmology the wavefunction of the universe is obtained using the ADM formalism and canonical quantization that lead to the Wheeler-deWitt equation \cite{dewitt1967}. Combining this with de Broglie's guidance equations, we can obtain quantum trajectories for the gravitational and matter degrees of freedom that can describe the quantum regime, and they usually connect naturally with the classical evolution on large length scales through the dynamical equations. 

The trajectories for a quantum bounce driven by a scalar field with only the kinetic term were obtained first in \cite{Pinto-Neto2000} and more recently in \cite{colinpinto-neto2017} the full equations including the exponential potential were also solved demonstrating non-singular behavior and appropriate classical limits before and after the bounce. In all the models we will study in this paper the bounce itself is dominated by the kinetic term of the scalar field. So, we briefly review below the canonical quantization of the FRLW inflationary model and how we can use Pilot-Wave Theory to calculate quantum trajectories. 

Using the ADM formalism, when the dynamics is dominated by the kinetic term, we can write the Hamiltonian for a scalar field in a Friedmann universe as 
\begin{equation}
    H = N \mathcal{H} = \frac{N \kappa^2}{12 \mathcal{V} e^{3\alpha}}(-\Pi^2_{\alpha} + \Pi^2_ {\phi}), 
\end{equation}
where $N$ is the lapse function, $\mathcal{V}$ is the volume of the conformal hypersuface and $\Pi_i$ are the canonical momenta associated with each variable. For the purposes of this section we use the dimensionless scalar field $\kappa \phi / \sqrt{6}$. The classical momenta are given by 
\begin{equation}
    \Pi_{\alpha} = - \frac{6\mathcal{V}}{N\kappa^2}e^{3\alpha}\dot{\alpha}, \ \ \ \ \ \ \ \  \Pi_{\phi} = \frac{6\mathcal{V}}{N\kappa^2}e^{3\alpha}\dot{\phi},
\end{equation}
and the conformal hypersurface volume was defined as $\mathcal{V} = 4\pi l_P^3/3$ (where $l_p \equiv \sqrt{G_N}$) in order to make the Universe reach the Planck volume when $a=1$. 

Applying the Dirac quantization method and imposing the Hamiltonian constraint on the quantum state of our model we obtain the Wheeler-DeWitt equation as
\begin{equation}
    \hat{\mathcal{H}}\Psi(\alpha, \phi) = 0 \Rightarrow \left[ -\frac{\partial^2}{\partial \alpha^2} + \frac{\partial^2}{\partial \phi^2} \right]\Psi(\alpha, \phi) = 0.
\end{equation}
It has a general solution given by 
\begin{multline}
     \Psi(\alpha, \phi) = F(\phi + \alpha) + G(\phi - \alpha) \\
    \equiv \int dk [ f(k)e^{i k (\phi + \alpha)} + g(k)e^{i k (\phi - \alpha)}],
\end{multline}
where $f$ and $g$ are arbitrary functions. 

The first step in the Pilot-Wave theory is to obtain the guidance equations which can be read from the Hamilton-Jacobi-like equation one obtains when the wave functions is expressed in its polar form. Making $\hbar = 1$ and writing $\Psi=Re^{iS}$, from equation (20) we can derive
\begin{equation}
    \left(\frac{\partial S}{\partial \alpha}\right)^2 - \left(\frac{\partial S}{\partial \phi}\right)^2 - \frac{1}{R}\left(\frac{\partial^2 R}{\partial \alpha^2} -\frac{\partial^2 R}{\partial \phi^2}  \right) = 0. 
\end{equation}
The last term depends exclusively on the amplitude of the wave function $R$ and is referred to as the ``quantum potential''\footnote{This nomenclature is misleading since $Psi$ is not a field and does not produce a force in the particles through a potential in the standard sense. Even when one interprets the wave function as some type of \textit{real} entity it cannot be a field since it evolves in configuration space in a way that is fundamentally different from standard quantum fields.} and when it becomes negligible the above equation reduces to the classical Hamilton-Jacobi. Assuming $\alpha$ and $\phi$ to have well defined trajectories at all times, the guidance equations will be given by 
\begin{equation}
      \Pi_{\alpha} = \frac{\partial S}{\partial \alpha} =- l_Pe^{3\alpha}\dot{\alpha}, \ \ \ \ \ \ \ \  \Pi_{\phi} = \frac{\partial S}{\partial \phi}= l_Pe^{3\alpha}\dot{\phi},
\end{equation}
where we also made $N=1$. 

Different prescriptions for functions $f$ and $g$ will generate a variety of possible dynamics for the bounce, some of which were explored in \cite{colinpinto-neto2017}. As an example, we can use a Gaussian superposition of plane waves with the definition \cite{Pinto-Neto2000}
\begin{equation}
    f(k) = g(k) = exp\left[ \frac{-(k-d)^2}{\sigma^2}\right]
\end{equation}
to obtain a specific phase $S(\alpha, \phi)$ that we can substitute in the guidance equations obtaining 
\begin{align}
\label{system-bounce}
    l_P \dot{\alpha} =\frac{\phi \sigma^2 \sin(2 d \alpha ) + 2 d \sinh (\sigma^2 \alpha \phi)}{2e^{3\alpha} [\cos (2d\alpha)+\cosh(\sigma^2\alpha\phi)]}, \nonumber \\
    lp\dot{\phi} = \frac{-\alpha \sigma^2 \sin (2d\alpha)+2d\cos(2d\alpha) +2d \cosh(\sigma^2 \alpha \phi) }{2e^{3\alpha}[\cos(2d\alpha) + \cosh (\sigma^2 \alpha \phi)]}.
\end{align}

\begin{figure}[H]
\centering
\includegraphics[scale=0.45]{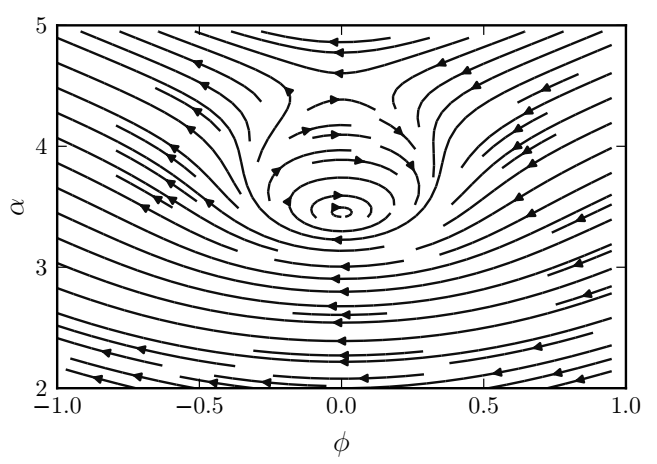} 
    \caption{Phase space trajectories for the system of equations \ref{system-bounce}, when $d = -1$ and $\sigma = 1$. The Figure depicts both bouncing and cyclic trajectories. }
    \label{fig:phasespace_alphaphi}
\end{figure}

Note that at large length scales, $\alpha \gg 1$, the system \ref{system-bounce} naturally implies that ${\rm d} \alpha / {\rm d} \phi \approx \pm 1$, which is the classical limit of stiff matter domination. It is at these points where we connect the classical solutions discussed in Ref.~\cite{Heard_2002}. One can see from Figure \ref{fig:phasespace_alphaphi} that the bounce necessarily connects the points $A_+$ to $A_-$, and vice-versa.


The parameters $d$ and $\sigma$ define the characteristics of the wavepacket, and hence are directly related to the quantum nature of the bounce. 

%% file: sections/section04.tex
\section{2D Phase Space and Dynamical Evolution Analysis} \label{sec:phasespace}

In this section we will discuss the consequences of introducing matter fluids in the model discussed so far that included only the scalar field with an exponential potential. Our numerical method sets initial conditions for the fluid variable $z$ and the potential variable $y$ fixed at $10^{-5}$ for all our plots in order to avoid instabilities that happen when $x$ gets too close to $\pm1$. The initial values do affect the overall evolution and it will be important in future works to get a more precise description of the transition from the quantum to the classical phase if we are interested in investigating the evolution of perturbations and possible quantum signatures. However, to study the dynamics of the background it suffices to assume that right after (before) the bounce some of the kinetic energy of the scalar field is transferred to (from) its potential and some type of perfect fluid. We used Maple software to numerically solve equations (14), (15) and (16) simultaneously using the Runge-Kutta-Fahlberg numerical method of order 4 with relative and absolute error tolerances of $10^{-20}$. We chose different values of the constant $\gamma$ (remembering that $w = \gamma - 1$ represents the EoS parameter of the fluid) and observed what different dynamics could be obtained when we varied the field parameter $\lambda$. In the next three subsections we show our results for the case where $\lambda = \sqrt3$, that was the case studied in previous works \cite{colinpinto-neto2017, Bacalhau2018, Frion2024bounce}. In the last subsection we focus on an interesting case with $\lambda=2$ that has some interesting potential implications for more realistic models. 

We would like to emphasize that although the bouncing solutions we present are asymmetric in time, all models can be time-reversed because the original classical Friedmann equations are time-reversible invariant, and the quantum bounce connection can be reverted from $A_-$ ($A_+$) to $A_+$ ($A_-$). We present only the most suitable models to be potentially compatible with observations.

\subsection{Dust scalar field plus a cosmological constant}

We start by analyzing the representative case where the matter fluid is a cosmological constant. We take the scalar field with $\lambda = \sqrt{3}$, with dust attractor (repeller) behavior, but the results are independent of $\lambda^2 < 6$.


\begin{figure}[H]
\centering
\includegraphics[scale=0.45]{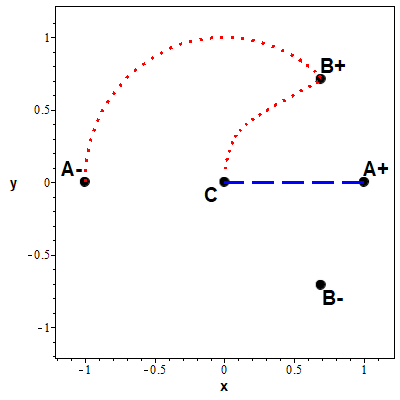} 
    \caption{Phase space trajectory for the case with $\lambda=\sqrt{3}$ and $w=-1$ (dark energy). Point
C is the attractor (repeller) of the expanding (contracting)
phase where the matter fluid dominates. In this case point B+ is an unstable attractor of
the expanding phase. The contraction goes from $C$ to $A+$ (dashed blue line) and the expansion from $A-$ to $C$ passing through $B+$ (red dotted line).}
    \label{fig:phasespace2}
\end{figure}

\begin{figure}[H]
\centering
\begin{tabular}{c}
\includegraphics[width=.5\textwidth, height=1.2in]{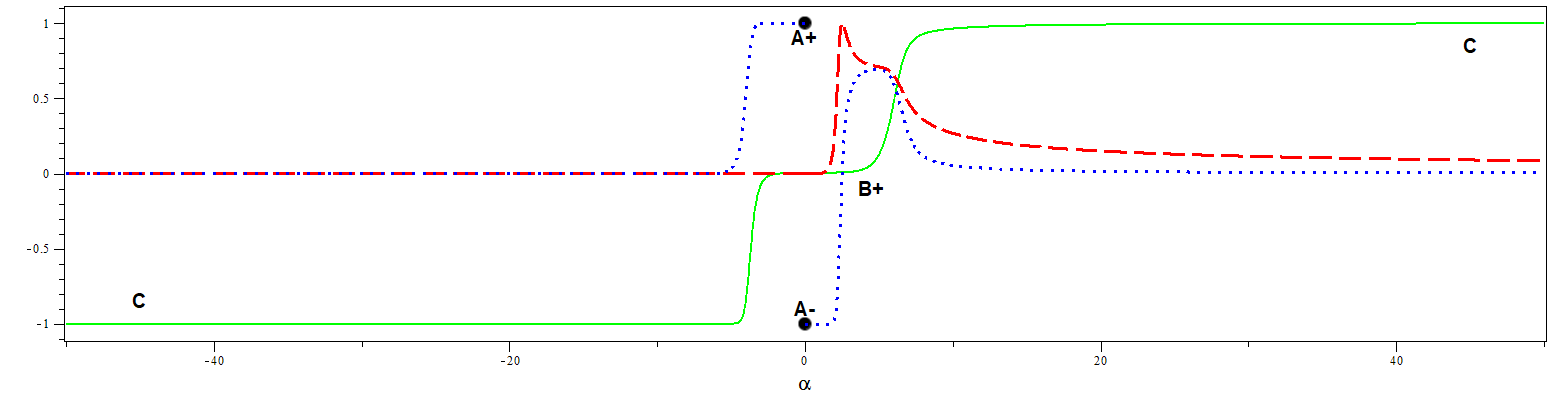}  \\(a)  \\
  \includegraphics[width=.5\textwidth, height=1.2in]{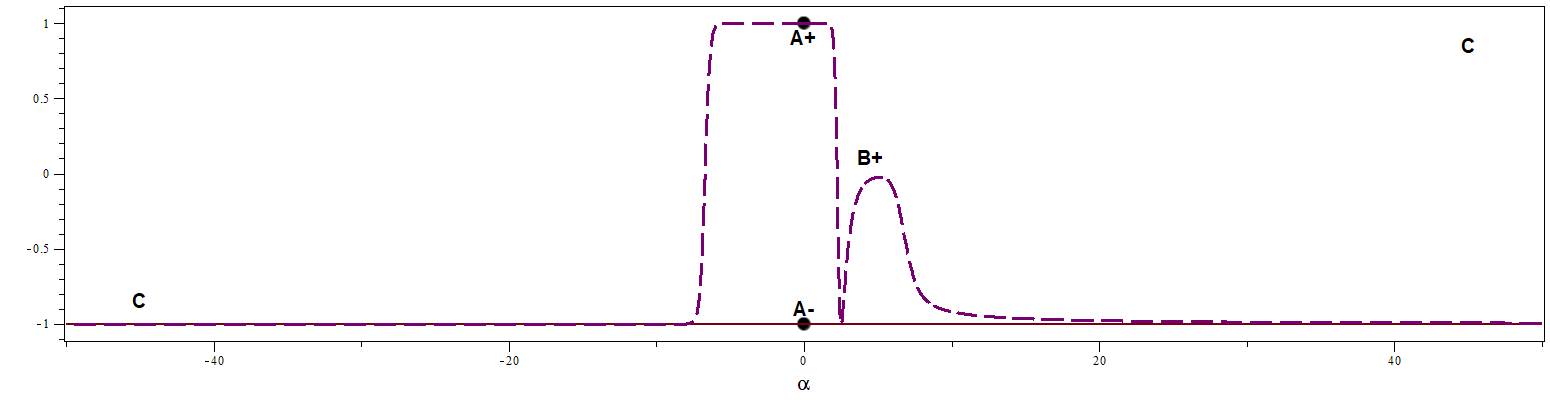} \\
(b)  \\
\end{tabular}
\caption{(a) Evolution of the dynamical system variables $x(\alpha)$ (dotted blue line), $y(\alpha)$ (dashed red line) and $z(\alpha)$ (solid green line). (b) EoS parameters for the matter fluid (in this case, $w=-1$, solid line) and the scalar field $w_{\phi}$ (dashed line).}
\label{fig:xywevol2}
\end{figure}

As expected, the cosmological constant spoils the initial dust dominated state of the contraction as we can see in Figure \ref{fig:xywevol2}. 

In Figure \ref{fig:phasespace2} we plot the evolution in the $(x, y)$ phase space. The blue dashed line represents the contraction phase and the red dotted line represents the expanding phase. The evolution starts near the point $(0, 0)$ where the fluid dominates and goes to $(1,0)$ near the bounce. The quantum bounce moves the solution from $(1,0)$ to $(-1,0)$, which goes from $(-1,0)$ to $(0,1)$ before decaying to the kinetic-potential scaling behavior, and then to the fluid dominated attractor. It is important to point out that the asymptotic solution has $y \neq 0$.

In Figure \ref{fig:xywevol2} we plot the evolution of the transformed variables representing the kinetic energy density of the scalar field ($x$ blue dotted), the potential energy density ($y$ red dashed) and the density of the "matter" fluid ($z$ green solid). In Figure 4b we see the effective EoS parameter for the scalar field ($\omega_{\phi}$ purple) and the EoS for the cosmological constant/dark energy fluid ($w=-1$). The universe starts dominated by the cosmological constant. The scalar field is insignificant in density and has an effective equation of state given by $P_{\phi} = - \rho_{\phi}$. As the universe collapses it passes through a stiff matter phase dominated by the scalar field. After the bounce the scalar field potential dominates as $\omega_{\phi} \rightarrow -1$, and then transits to a brief potential-kinetic scaling period with a dust-like behavior ($\omega_{\phi} \approx 0$). When the cosmological constant/dark-energy fluid starts to dominate, the scalar field tracks the fluid and goes back to an effective dark-energy behavior with the potential density asymptotically going to a constant non-vanishing value. 

The time-reversal solution would have a period of dust domination during the contracting phase after the cosmological constant/dark-energy fluid domination. However, the expanding phase would be completely dominated by the cosmological constant/dark-energy fluid just after the scalar field kinetic energy domination, which is not realistic.

This type of model is physically unreliable: the presence of the cosmological constant/dark-energy fluid spoils the nice properties of the simpler model containing only the scalar field discussed in Ref.~\cite{Bacalhau2018} and summarized in sub-section II-A, which is physically reliable and simpler. However, this example serves to illustrate both how the presence of standard dark energy might spoil the desired initial dust-like conditions, and how it might be possible to have an accelerated phase followed by a matter dominated phase after the bounce even in the presence of $\Lambda$. At late times, although the dark energy fluid dominates, there is still a non-vanishing contribution of the scalar field with $w_{\phi}$. This could signal - and we will verify that it does in the following figures - that the scaling behavior of the scalar field could allow it to survive until later times and play a role in the cosmological evolution as a non-standard matter fluid (like dark energy or dark matter).

\subsection{Dust scalar field plus dust fluid}

We turn to the case where the matter fluid is modeled as dust with $w=0$ ($\gamma = 1$), and the scalar field still have $\lambda = \sqrt{3}$. The phase space trajectories in Figure 5 show the evolution starting near $(0.7, -0.9)$. The expanding phase goes near the $(0,0)$ fluid-dominated point and then to an approximate scaling kinetic-potential solution.
\begin{figure}[H]
\centering
\includegraphics[scale=0.45]{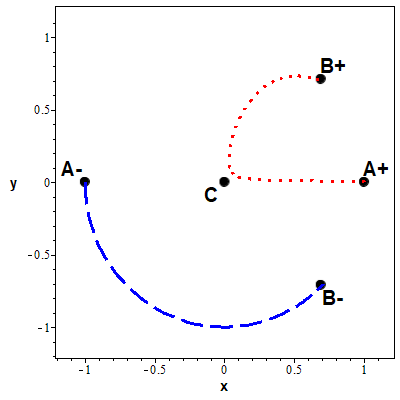} 
    \caption{Scalar field with potential $V = \exp{-\sqrt{3}\kappa \phi}$ and $w=0$ (dust). $B-$ is the initial repeller that drives the contracting phase to $A-$ (dashed blue line). After the quantum bounce and the initial stiff matter phase $A+$ there is a period of matter domination ($C$). The final attractor is $B+$ but the fluid component has a non-vanishing contribution at late times (red dashed line).}
    \label{fig:phasespace3}
\end{figure}

\begin{figure}[H]
\centering 
\begin{tabular}{c}
\includegraphics[width=.5\textwidth, height=1.2in]{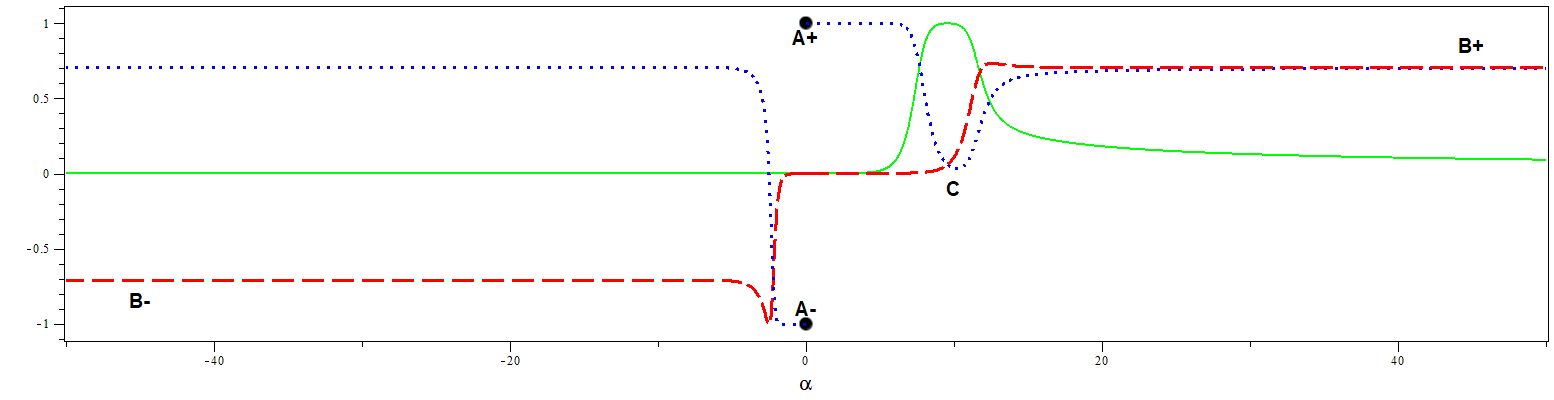}  \\(a) \\
  \includegraphics[width=.5\textwidth, height=1.2in]{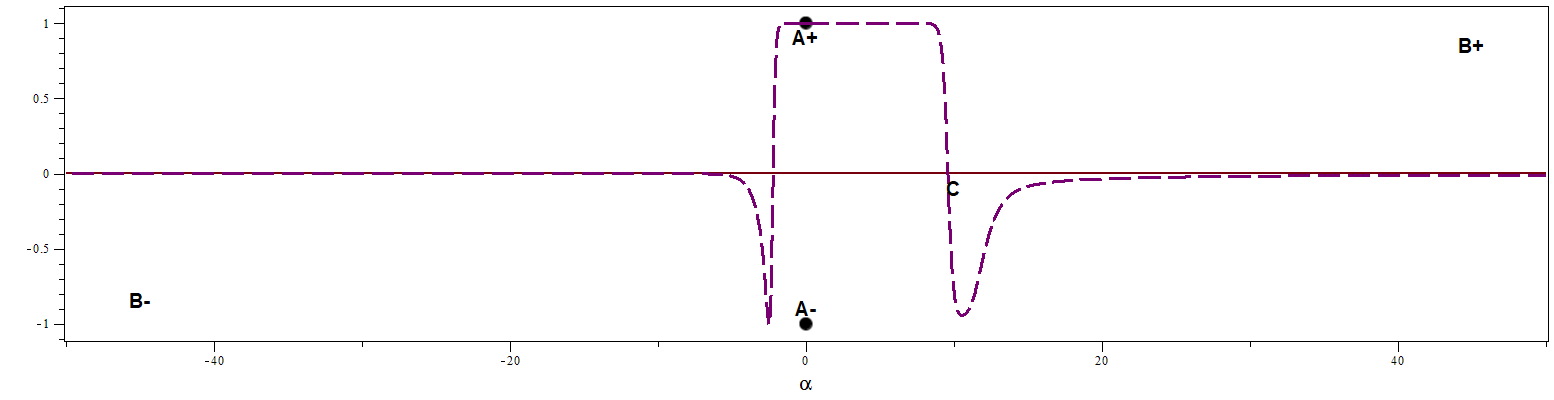} \\
(b)  \\
\end{tabular}
\caption{(a) Evolution of the dynamical system variables $x(\alpha)$ (dotted blue line), $y(\alpha)$ (dashed red line) and $z(\alpha)$ (solid green line). (b) EoS parameters for the dust matter fluid ($w=0$, solid line) and the scalar field $w_{\phi}$ (dashed line). Potential and fluid densities near the bounce are set as $y_0 = 10^{-5}, z_0 = 10^{-5}$. The kinetic term is $x \approx 1$ before and $x \approx -1$ after the bounce.}
\label{fig:xywevol3}
\end{figure}

In Figure \ref{fig:xywevol3} the evolution starts near the scaling kinetic-potential solution with the scalar field behaving like dust. The fluid is unimportant during the whole contracting phase. Before reaching the bounce, the scalar field goes through a dark-energy like phase. The kinetic term dominates around the bounce. Afterwards, the model passes through a matter-fluid dominated phase. Meanwhile, the scalar field goes through a phase where $\omega_{\phi} \approx -0.9$. The scalar field goes back to the potential-kinetic solution with dust behavior and starts to dominate as the matter fluid decays to a small but non-vanishing density.

This model also provides some interesting features. The contraction is totally dominated by the dust-scaling behavior of the scalar field, leading to a scale invariant spectrum of cosmological perturbations. However, before the bounce, the scalar field passes through a dark energy phase, which may slightly modify the spectrum index of these perturbations.

After the bounce, the dust fluid dominates for a while. In this phase the field seems to play no role, but its transition from $w_\phi = 1$ to $w_\phi \approx -0.9$ and then back to zero could have non-trivial effects in the phase of matter formation. Note that in the expanding phase where $w_\phi \approx -0.9$ the matter fluid dominates, but not totally. As the dust contraction can already yield an almost scale invariant spectrum of scalar cosmological perturbations, it is not necessary that the $w_\phi \approx -0.9$ behavior be long and strong enough to yield sufficient inflation. It can be very short, but sufficient to inject some energy to the expansion rate and potentially lead to a small increase of the today's Hubble radius as measured by early universe observations, alleviating the Hubble tension.
Another thing to notice in this model is that the matter fluid still contributes in the far future with a small density. This feature may be used to model the baryonic matter by the dust fluid, which becomes relevant only after the bounce, and the scalar field to model dark matter. This possibility will be further explored in the next examples.

The time reversed model of this example is physically uninteresting.

In the above examples, there is no radiation phase. We will explore this possibility in the next examples.

\subsection{Dust scalar field plus a radiation fluid}

\begin{figure}[H]
\centering
\includegraphics[scale=0.45]{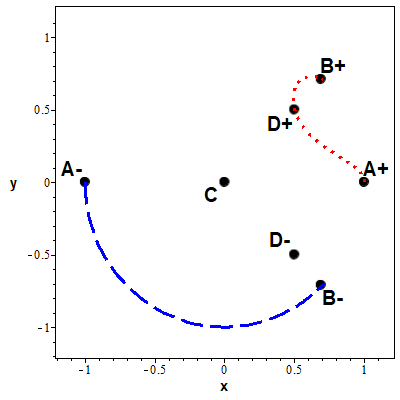} 
    \caption{Trajectory for a model with $\lambda = \sqrt{3}$ and $w=1/3$ (radiation). $B-$ (where $w_{\phi}=0$) is the initial repeller that drives the contracting phase to $A-$ (blue dashed line). After the quantum bounce and the initial stiff matter phase $A+$ there is a period of scaling behavior ($D+$). The final attractor is $B+$ (red dotted line).}
    \label{fig:phasespace4}
\end{figure}

\begin{figure}[H]
\centering
\begin{tabular}{c}
\includegraphics[width=.5\textwidth, height=1.2in]{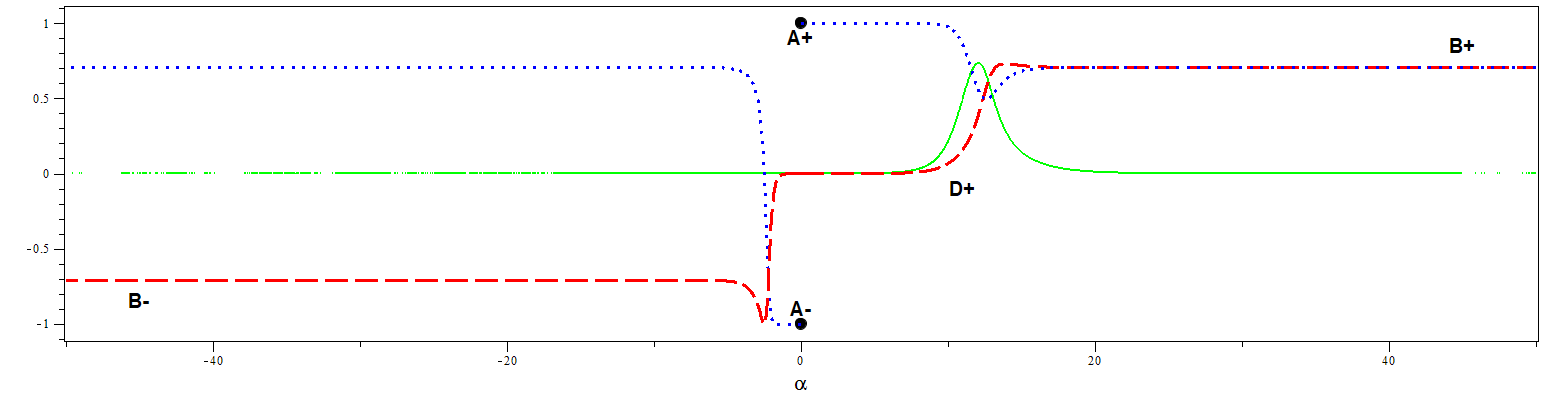}  \\(a) \\
  \includegraphics[width=.5\textwidth, height=1.2in]{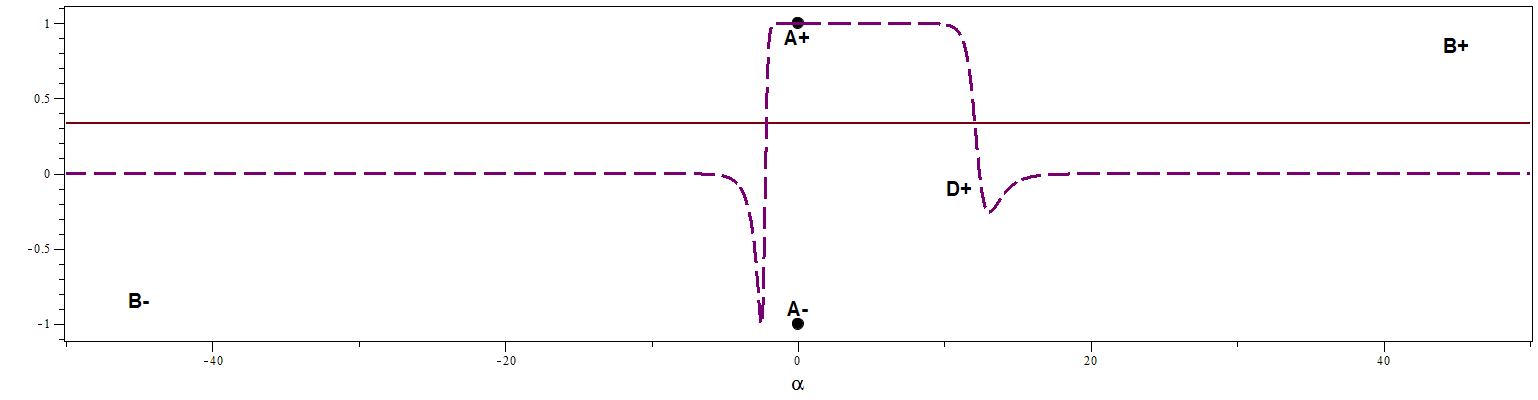} \\
(b) \\
\end{tabular}
\caption{(a) Evolution of the dynamical system variables $x(\alpha)$ (dotted blue line), $y(\alpha)$ (dashed red line) and $z(\alpha)$ (solid green line). Scalar field with potential $V = \exp{-\sqrt{3}\kappa \phi}$ and $w = 1/3$. (b) EoS parameters for the radiation fluid ($w=1/3$, solid line) and the scalar field $w_{\phi}$ (dashed line). Potential and fluid densities near the bounce are set as $y_0 = 10^{-5}, z_0 = 10^{-5}$. The kinetic term is $x \approx 1$ before and $x \approx -1$ after the bounce.}
\label{fig:xywevol4}
\end{figure}

We now introduce a radiation fluid with $w=1/3$ ($\gamma = 4/3$), keeping the dust scaling scalar field. In Figures \ref{fig:phasespace4} and \ref{fig:xywevol4} we see that the dust scaling phase of the scalar field dominates either the asymptotically initial contracting era and the final expanding era, passing in between through an stiff matter quantum bounce. Right before the bounce there is a brief period of potential domination when $\omega_{\phi}$ passes through $\approx -1$. There is also a period during expansion when the radiation fluid contributes significantly to the total density.

The radiation fluid becomes important only in the expanding phase, when it competes with the scalar field at early stages, and then becomes less relevant as the Universe expands, as usual. Again, during the phase of both radiation fluid and scalar field relevance, the scalar field presents a $w_\phi$ small but negative, which may also inject some energy to the expansion rate, yielding a small increase of the today's Hubble radius as measured by early universe observations, and also alleviating the Hubble tension.

The time reversed model of this example is physically uninteresting.


\subsection{Radiation scalar field plus a dust fluid} 
\label{sec:dust-to-dust}

By surveying the space of solutions given by different values of the potential parameter $\lambda$, we have found a model that has appropriate dust-like initial conditions and an interesting dynamics that can model the early stages of the expansion phase. During the contraction the model is always dominated by a dust-like fluid composed of both the scalar field and the matter fluid contributing to the dynamics. This is the scalar field-fluid tracking that was described in \cite{Heard_2002}.

The model consists of a radiation scaling scalar field with $\lambda = 2$, implying $w_{\phi}=1/3$ at the attractor (repeller) points of the scalar field evolution. It however tracks the dust fluid at both asymptotic limits, as it can be seen in Figure \ref{fig:xywevol5}. Near the bounce the dynamical evolution is completely dominated by the scalar field kinetic energy (stiff behavior), and we can use the quantum bounce solutions of Ref.~\cite{Pinto-Neto2000} to connect the contracting phase to the expanding phase. 


\begin{figure}[H]
\centering
\includegraphics[scale=0.45]{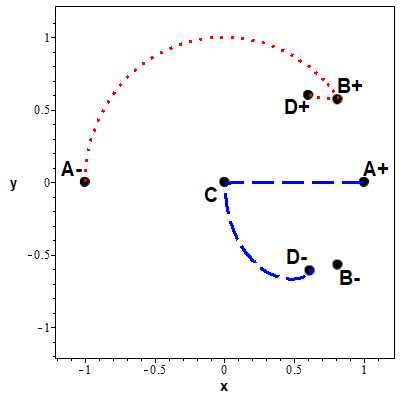} 
    \caption{The scalar field now has parameter $\lambda=2$ and $w = 0$ for the matter fluid. The contracting phase starts at $D-$ and goes through $C$ before reaching the bounce ($A+$) (blue dashed line). In the expanding phase (red dotted line), the universe goes through an initial accelerated expansion ($w_{\phi} = -1$ when $y = 1$), has a field dominated phase ($B+$, where $w_{\phi} = 1/3$) and than goes to the $D+$ fluid potential-kinetic scaling solution with $w_{\phi} = 0$.}
    \label{fig:phasespace5}
\end{figure}

At the beginning of the expanding phase the presence of the dust fluid is  irrelevant, and we have an inflation-like period dominated by the scalar field potential which transitates to its scaling behavior that causes the EoS parameter to become
\begin{equation}
    w_{\phi} = 1/3,
\end{equation}
for some period of time. It is possible that this scenario can describe the graceful exit from inflation into the radiation dominated era with a single field.

Another interesting feature happens with the asymptotic behavior of the solutions. The scalar field tracks the dust fluid behavior and becomes a dust-like scalar field, contributing with 
\begin{equation}
    x^2 + y^2 \approx 2 \cross (0.6123)^2 \approx 0.75
\end{equation}
to the total matter content of the model. 
This means that the total asymptotic matter content of this universe is divided between normal matter (baryons?) making up $25\%$ of the total energy density, and a scalar field behaving as dust dominating the universe with $75\%$ (dark matter?). 

\begin{figure}[H]
\begin{tabular}{c}
\includegraphics[width=.5\textwidth, height=1.2in]{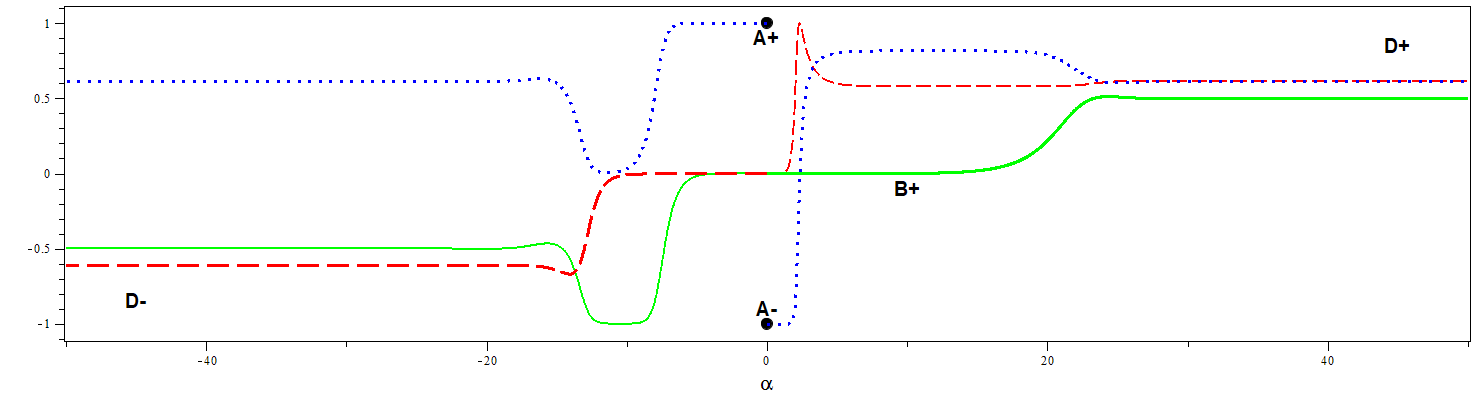}  \\(a)  \\
  \includegraphics[width=.5\textwidth, height=1.2in]{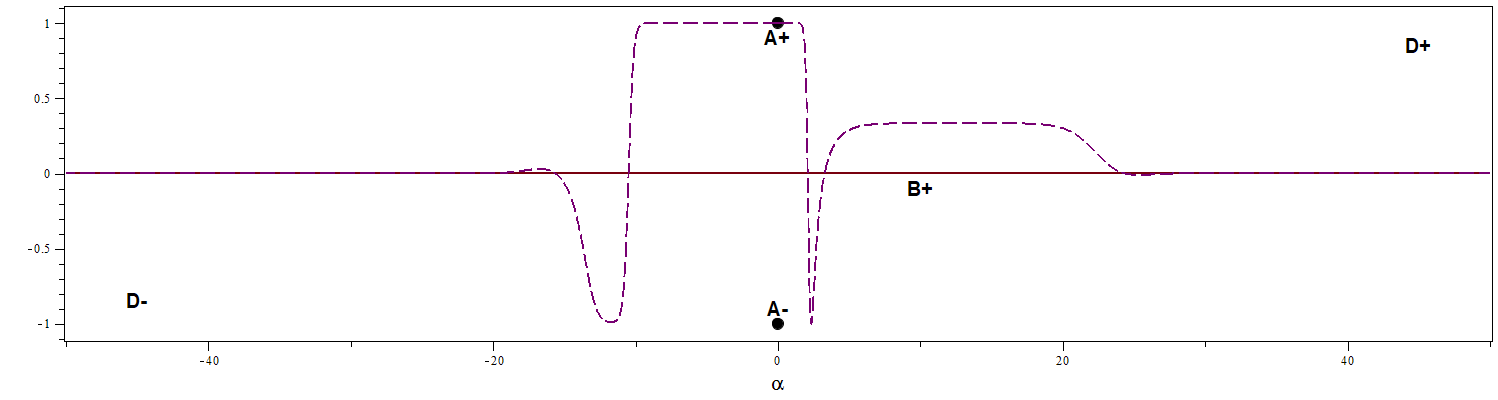} \\
(b) \\
\end{tabular}
\caption{(a) Evolution of the dynamical system variables $x(\alpha)$ (dotted blue line), $y(\alpha)$ (dashed red line) and $z(\alpha)$ (solid green line). (b) EoS parameters for the dust matter fluid $w=0$ (solid line) and the scalar field $w_{\phi}$ (dashed line). Scalar field with potential $V = \exp{-2\kappa \phi}$ ($\lambda = 2)$.}
\label{fig:xywevol5}
\end{figure}

Hence we have found a model (Figures \ref{fig:phasespace5} and \ref{fig:xywevol5})  where its expanding phase after the bounce contains a primordial inflationary era followed by the scaling radiation phase of the scalar field
which then tracks the dust fluid yielding a dust universe composed partially of `normal' dust matter, maybe baryons, and a scalar field dust-like matter fluid, maybe dark matter. Note also that if the inflationary phase lasts long enough, it is not necessary to have a long contracting dust dominated phase in order to have an almost scale invariant spectrum of cosmological perturbations.

The time reversed model of this example is physically uninteresting.

%% file: sections/conclusions.tex
\section{Discussion} \label{sec:discussion}

Scalar fields and tracking solutions have long been used to model cosmological scenarios where inflation and the present dark energy phase could both be driven by the same scalar field \cite{Sami2022}. Besides that, there are several modifications of General Relativity motivated by the cosmological constant problem that can be effectively described by scalar field-like fluids with different, possibly exotic, potentials. 
Our analysis shows that the introduction of simple barotropic fluids in the classical contracting and expanding phases in scalar field scenarios generates new classes of tracking bouncing models which deserve to be studied.


The matter content of the universe today is believed to be made up of $\sim 85\%$ dark matter and $\sim 15\%$ baryonic matter. Considering our scalar field as an effective model of some more complete field theory, it is possible to envision an universe where dark matter could actually be made of scalar field weakly interacting particles. The possibility of scalar fields generating dark matter, although apparently less promising than using it to explain dark energy, is not ruled out by experiments \cite{Jesus_2016, Arakawa2021}. Furthermore, there are also proposals to modify General Relativity through an $f(R)$ function in the Einstein-Hilbert action, or using non-minimal couplings between matter and curvature, that lead to extra degrees of freedom that could be effectively described by scalar fields and explain the dark matter effects \cite{Choudhury2016}. 

A single canonical scalar field with a simple exponential potential in the context of General Relativity 
together with a quantum bounce can lead to a long dust dominated contracting phase, and a late time accelerated expanding phase making the role of dark energy phase. Such a universe would have scale invariant spectrum of cosmological perturbations, plus other interesting observational features \cite{Bacalhau2018, Frion2024bounce}. However, the model is too simple to account for radiation, baryons and dark matter also, hence it needs extra ingredients to deliver a realistic picture of our universe. This was the main purpose of this paper: to investigate the possibilities when an extra barotropic fluid is added to such models.


The main conclusions of our analysis are:
\begin{itemize}
    \item Examples (B,C,D) of Section 4 show that dark matter and baryonic matter could possibly be modeled by the fluid and the field with the present right proportions. Also, these scenarios may model the back-reaction of baryon creation at the bounce as long as the fluid representing the baryons have negligible anergy densities in the contracting phase, or disappear at high energies. 
    \item The case D of Section 4 provides an interesting model with both the long matter dominated contraction and an initial accelerated expansion. There is also an early expanding radiation dominated phase just after inflation, where the scalar field plays the role of radiation. It then tracks the dust fluid, yielding two different matter species at larger scales. If the inflationary phase lasts long enough, there is no need of a matter bounce contracting phase to yield a scale invariant spectrum of scalar perturbations;
    \item In cases (B,C) the early expanding phase has also the contribution of the scalar field with negative effective equation of state parameter, which may supply sufficient energy to increase the Hubble parameter at early times and alleviate the Hubble tension;
    \item We have shown that in the presence of a fluid where $w \geq 0$, the transient phases where $w_{\phi}= -1$ are accompanied by a growth in matter density that avoid an accelerated expansion. Hence the scalar field is not dominant, and it cannot play the role of dark energy in these cases. A simple proposal to mimic dark energy could be considering non-minimal couplings \cite{Barman_2024};
    \item If the goal is to find the simplest model that can both explain dark matter \textbf{and} dark energy with a single scalar field, one possible next step would be to consider higher order curvature terms or other $f(R)$ type modifications that could generate our current accelerated expansion;
    \item Connected to the last point, in these models the scalar field has dominant transient dark energy behavior only in the contracting phase. This may change the spectral index of scalar perturbations, yielding a desirable red-tilt. Note also that a passage through $w_\phi = -1$ may impose large instabilities in cosmological perturbations when only the scalar field is present. This is not the case when another fluid or field is present, also because the background will never behave exactly as de Sitter in this case.
\end{itemize}

The next step is to calculate the evolution of cosmological perturbations in the models presented in this paper. These requires some work because we have now two matter degrees of freedom, implying the appearance of entropy perturbations, and judicious vacuum initial conditions for both adiabatic and entropy perturbations are necessary. This is the subject of our forthcoming works.

\section*{Acknowledgments}

FBL acknowledges the support of CNPq of Brazil
under grant PCI-DC 300127/2024-3. NPN acknowledges the support of CNPq of Brazil under grant PQ-IB 310121/2021-3.